\documentclass[12pt]{article}

\usepackage{latexsym,amsmath,amssymb,theorem,epsfig}

\usepackage{graphicx}
\usepackage{graphicx,subfigure}
\usepackage{epstopdf}
\usepackage[body={17.5cm, 21cm},right=2cm]{geometry}
\usepackage{amssymb}
\usepackage{amsmath}
\usepackage{psfrag}
\usepackage{epsfig}
\usepackage{cancel}
 \allowdisplaybreaks[4]

\usepackage[all]{xy}

\DeclareFontFamily{OT1}{rsfs10}{}
\DeclareFontShape{OT1}{rsfs10}{m}{n}{ <-> rsfs10 }{}
\DeclareMathAlphabet{\mathscript}{OT1}{rsfs10}{m}{n}

\numberwithin{equation}{section}

\newcommand{\ns}{\normalsize}

\newcommand{\cH}{{\cal H}}

\newcommand{\be}{\beta}

\def\gsim{ \lower .75ex \hbox{$\sim$} \llap{\raise .27ex \hbox{$>$}} }
\def\lsim{ \lower .75ex \hbox{$\sim$} \llap{\raise .27ex \hbox{$<$}} }
\def\be{\begin{equation}}
\def\ee{\end{equation}}
\def\bea{\begin{eqnarray}}
\def\eea{\end{eqnarray}}

\begin{document}

\begin{titlepage}

\vspace{-5cm}

\title{
  \hfill{\ns }  \\[1em]
   {\LARGE  Modifying Gravity in the Infra-Red by imposing\\
    an ``Ultra-Strong" Equivalence Principle\footnote{Based on the essay written for the Gravity Research Foundation 2009 Awards.}.}
\\[1em] }
\author{
   Federico Piazza\footnote{fpiazza@perimeterinstitute.ca}
     \\
   {\ns Perimeter Institute for Theoretical Physics}\\
{\ns Waterloo, Ontario, N2L 2Y5, Canada}}

\date{}

\maketitle

\begin{abstract}
The equivalence principle suggests to consider gravity as an infra-red phenomenon, whose effects are visible only outside Einstein's free-falling elevator. By curving spacetime, General Relativity leaves the smallest systems free of classical gravitational effects.
However, according to the standard semi-classical treatment, indirect effects of gravity can be experienced inside the elevator through the well-known mechanism of quantum particle production. Here we try a different path than the one historically followed: rather than imposing field quantization on top of a curved manifold, we attempt to upgrade the equivalence principle and extend it to the quantum phenomena. Therefore, 
we consider, and try to realize in a theoretical framework, a stronger version of the equivalence principle, in which all the effects of gravity are definitely banned from the elevator and confined to the infra-red. For this purpose, we introduce infra-red modified commutation relations for the global field operators (Fourier modes) that allow to reabsorb the time-dependent quadratic divergence of the vacuum expectation value of the stress-energy tensor. The proposed modification is effective on length scales comparable to the inverse curvature and, therefore, does no add any dimensional parameter to the theory. 

\end{abstract}

\thispagestyle{empty}

\end{titlepage}

Modifications of General Relativity (GR) on the largest scales have been advocated in order to give account for the present acceleration of the Universe. These alternatives to GR look now particularly appealing  in view of some emerging tensions between standard $\Lambda$CDM cosmology and large-scale observations \cite{justin}.
Infra-red (IR) modifications of gravity typically involve large extra-dimensions and are effectively equivalent to giving a small mass to the graviton . 

In this note we explore a modification of GR of quite a different nature and contemplate the possibility that the very geometrical description of space-time as a metric manifold may break down on the largest scales. 
This point of view is provocative only in appearance; it aims in fact, rather conservatively, to recover the most genuine and intuitive physical content of the Equivalence Principle (EP), namely, the absence of any gravitational effect within each sufficiently small free-falling system. 

\section{Invitation: Gravity as an Infra-Red Effect.}\label{secinv}

The equivalence principle (EP) can be formulated simply as follows: \emph{inside a sufficiently small free-falling elevator you do not see the (classical) effects of gravity}. Amusingly, such a cornerstone of modern physics is actually stating what  (where) gravity is not, rather than what  (where) gravity is!  Among the many celebrated implications of general relativistic physics, the view that we aim to stress here is that EP forces us to consider and describe gravity as an IR phenomenon, whose effects are visible only outside the free-falling elevator. How EP turned into a consistent theory is well known: gravity is beautifully encoded in GR as the geometry of the physical space-time and therefore its effects are automatically suppressed within those systems that are much smaller than the inverse curvature. 
By changing (curving) the large-scale structure of spacetime, GR makes the smallest systems free of classical gravitational effects. Notably, the  IR scale where non-gravitational physics breaks down is not a parameter of the theory, but is set by the local curvature $R$. Schematically, in three dimensions, the area of a two-sphere of radius $l$ and volume $V$ receives corrections from flat-space expectation of the type
\begin{equation} \label{1}
A(l) \ = \ 4 \pi l^2 \, (1 + {\cal O}(l^2 R)) \ =\  (36 \pi V^2)^{1/3}\, (1 + {\cal O}(R V^{2/3}))  .
\end{equation}

The effects of gravity, originally banned from the free-falling elevator, reappeared, after the developments of quantum theory, through what one might call the back door.
The fields quantized on a curved manifold are sensitive to the global properties of spacetime because their modes are defined on the whole of it. As a result, inside the elevator, you will generically experience, and possibly detect with your local instruments, 
particle creation because of non-local gravitational effects. 
Clearly, the process of quantum particle creation does not contradict EP, which was formulated within the framework of classical physics. Nevertheless, it is tempting to try a different path than the one historically followed: rather than imposing field quantization on top of a curved manifold, here we attempt to upgrade the equivalence principle and extend it to the quantum phenomena.
Thus, we will consider a stronger version of EP, in which all the effects of gravity are definitely forbidden inside the elevator, including the quantum effects that in the standard semi-classical treatment lead to particle creation. 
More precisely, 

\begin{quote}
{\bf Equivalence Principle, ``Ultra-Strong" Version:}
For each matter field or sector sufficiently decoupled from all other matter fields, there exists a state, the ``vacuum", that is experienced as empty of particles by  each free-falling observer.
\end{quote}

Since the effect that we want to cancel is sensitive to the global structure of spacetime, we argue that applying the above more severe version of EP forces a further substantial change  at scales comparable to the curvature. Accordingly, we attempt to set up a theoretical framework for semi-classical gravity where the metric manifold structure of GR is systematically modified in the infra-red but holds in the vicinity of each point/event. An obvious warning is that, while EP is extremely well tested (see e.g. \cite{torsion}), the proposed ``ultra-strong" version is not. If any, experimental hints might actually be arguing against it, since, according to the current paradigm,  cosmological fluctuations are generated during inflation precisely with the mechanism of quantum particle creation. 
On the other hand, the appeal of the model that we are going to propose (see also \cite{newfedo}) is that it contains no more parameters than GR itself, and therefore it is in principle very well testable.

\section{Strategy}

Let us see what are the terms that have to be reabsorbed in order to realize the ``Ultra-Strong" equivalence principle. In the standard treatment,
the vacuum expectation value of the local energy density of a field can be expanded, at  high momenta, as follows
\begin{eqnarray} \label{structure}
\langle T_0^0\rangle_{\rm bare} &=& \int d^3 k\left(k  + \frac{f_{\rm quad}(t)}{k} + \frac{f_{\rm log}(t)}{k^3} + \dots  \right) \\[2mm]
& = &\ \  {\rm local\ terms}\ \ +\ \ {\rm non \ local\ terms}. \nonumber
\end{eqnarray}
Spatial homogeneity has been assumed for simplicity and the $f(t)$ are functions of time of appropriate dimension. The local terms account for the first, the second and part of the following pieces of \eqref{structure} and can be subtracted by local gravitational counterterms (cosmological constant, Planck mass etc\dots). The finite non-local terms represent the genuine particle content of the choosen ``vacuum" state. In order to impose the ``Ultra-Strong" equivalence principle one can attempt, in principle, two different strategies:
\begin{itemize}
\item Try to cancel the non-local terms \emph{after} the stress energy tensor has been renormalized by  usual means.
\item Circumvent the usual procedure of stress-tensor renormalization and just try to cancel the \emph{bare} time-dependent terms (every term in \eqref{structure} except the first) altogether.
 \end{itemize}
 
The second strategy is more radical but much more compelling for different reasons. First, the time-dependent pieces can arguably be reabsorbed with some IR modification, because they are effectively IR with respect to the quartically divergent term.
Second, bypassing the usual procedure of stress tensor renormalization might, at the same time, shed some new light on the cosmological constant problem, at least at the level of non-quantized semi-classical gravity. If, in the IR-modified theory,
all the time dependent terms just do not exist, then we do not need to renormalize the stress tensor anymore. Of course, we are still left with the initial quartic divergence, but we can live with it and treat it, as we do in flat space, by normal ordering. In standard semi-classical gravity, it is precisely the spacetime dependence of \eqref{structure} that makes the simple normal ordering prescription impossible.  Finally, it is interesting to reabsorb the  quadratic divergence in the IR, rather than with a local counterterm, because the required modification is of the right order of magnitude to give interesting cosmological implications \cite{newfedo}.

\section{The Idea: Regions of Space as Quantum Subsystems}

Before describing the model, it is worth examining closely what happens in the standard formulation of semi-classical gravity, where fields are quantized on a curved background manifold. Consider a spacetime with a global time foliation labeled by a time parameter $t$. The Universe as a whole at time $t$ is a three-dimensional manifold ${\cal M}$ in the GR description. The matter quantum fields are instead described by a quantum state living in a Hilbert space $\cH$. Thus, if now we consider a region of space (at time $t$) of finite volume $V$ (``this room, now"), that has two complementary descriptions \cite{fedo1,fabio1,sergio}: it is a \emph{submanifold} according to GR and a \emph{quantum subsystem} for the quantized fields. 
The correspondence sub-manifold/sub-system is explicitly realized by  the set of local operators $A(t, x)$ of the field theory. Those act on the quantum system $\cH$ but have labels $x$ living on the three-dimensional manifold ${\cal M}$. As a consequence, we can take integrals of scalar local operators over some region of volume $V$,
\begin{equation} \label{bah}
A(t, V) = \int_V d^3 x \sqrt{-g}\ A(t, x) ,
\end{equation}
which are still operators acting on the total Hilbert space $\cH$. More precisely, $A(t,V)$  acts non-trivially only on the quantum subsystem $\cH_V$ corresponding to the region of space that has been integrated over, and as the identity on the rest of the system $\cH_{\rm Rest}$. In this way, the algebra of operators $A(t, V)$ determines \cite{paolo} the partition $\cH = \cH_V \otimes \cH_{\rm Rest}$ of the quantum system/Universe and therefore can be taken as a definition of the region of space $V$. Defining regions of space as quantum subsystems \cite{fedo1} looks like a useless complication in the standard formalism because the correspondence  sub-system/sub-manifold is always implicitly at work: such a correspondence is set once and for all by \eqref{bah}, given the set of local operators $A(t,x)$.

When we say that the metric-manifold description might break-down, we mean, more precisely, that regions of space, still perfectly defined as quantum subsystems, may not have the nice property that the corresponding operators integrate as in \eqref{bah}.
We argue therefore that \eqref{bah} is valid only in the limit of zero curvature and that operators corresponding to extended regions of space can be written as \eqref{bah} only up to order ${\cal O}(R V^{2/3})$, where $R$ is some curvature scalar. Here is the mnemonic thumb-rule that will guide us in our further developments:  
\begin{equation} \label{2}
A(t, V) \ \simeq \  \int_V d^3 x \sqrt{-g}\ A(t, x) \ \left[1 + {\cal O}(R V^{2/3})\right] .
\end{equation}
We are deliberately mimicking the general type of corrections \eqref{1} that non-extensive geometrical quantities undergo in the transition from flat space to curved space. The idea is to extend this type of behavior also to extensive quantities, that in the standard description are just proportional to the volume. The implications of \eqref{2} are quite striking. Consider, for instance, a
perfectly homogeneous Universe. By definition, each comoving observer  measures in its surrounding the same energy density $\rho$. Eq. \eqref{2} implies that in that Universe, if one starts considering regions of space of Hubble size, the total energy 
inside that region will drastically differ from the three-dimensional integral of the local densities measured by the observers living therein.
In a metric manifold, that would be a dramatically non-local effect.

\section{Model Building: a Toy-Universe} \label{sec4}

We proceed in our construction by considering a compact, flat FRW Universe of total size $2 \pi a(t) L \ll H^{-1}$ much smaller than its own Hubble radius. According to our assumptions, as long as that condition is satisfied, the departure from usual semi-classical gravity are extremely small: this space-time, even globally, can be described as a manifold to a very good approximation. We will not touch the local QFT quantities but we will allow the globally defined operators (e.g. the ``Fourier modes") to  get ${\cal O}(L H)^2$ corrections. 
We follow a comoving observer/trajectory in this Universe and call $t$ its proper time.  For brevity, when we write $\vec{x} \approx 0$, we refer to a region of space around the trajectory considered small enough that we can define spatial derivatives at a point and introduce the usual local commutation relations between local fields and conjugate momenta. We then consider a massless scalar quantum field $\phi (t, \vec{x} \approx0)$ at the point 0 and strictly preserve its local dynamics as we know it. Our starting points are therefore the 
field equations in the Heisenberg picture along the comoving trajectory,
\begin{equation} \label{evo}
\ddot \phi(t, \vec{x} \approx 0) + 3 H \dot \phi (t, \vec{x} \approx 0) - \nabla^2  \phi (t, \vec{x} \approx 0)= 0
\end{equation}
and the Hamiltonian density ${\cal H}(t, \vec{x}\approx 0)$ at point 0, which is an explicitely time dependent operator:
\begin{equation} \label{hamtime}
T^0_0 =  {\cal H} = \frac{1}{2}\left(a(t)^3 \, \dot{\phi}^2 (t, \vec{x} \approx 0) + a(t)\, \vec{\nabla} \phi^2 (t, \vec{x} \approx 0) \right) .
\end{equation}
We can take the above equations as the definition of the local expansion $a(t)$. 

We now define global operators and make connections with the local ones defined at $\vec{x}\approx 0$. By definition, the Fourier modes  $\phi_{\vec{k}}$ of the field satisfy
\begin{equation} \label{fourier}
\phi(t, \vec{x}\approx 0) = \frac{1}{(2 \pi L)^3} \sum_{\vec{k}} \phi_{\vec{k}} (t) \ e^{i \vec{k} \cdot \vec{x}}.
\end{equation}
Again, at the exponent on the RHS the coordinates should be considered to extend as far as we need to define the derivatives of the field. Note that the coordinate $\vec{x}$ and momenta $\vec{k}$ are comoving with respect to the local expansion $a(t)$: $\vec{x} = \vec{x}_{\rm phys} / a(t)$, $\vec{k} = a(t) \vec{k}_{\rm phys}$ .
Since we are in a compact space, we expect  ${\vec{k}}$ to take discrete values and,  at least at zeroth order in our approximation, we have $k_i \simeq n_i/L$, the components $n_i$ being integer entries. By \eqref{fourier}, ${\vec{k}}$ is still the derivative at 0 in Fourier space: 
$\partial_i \ = i k_i$.

Global creators and annihilators are defined from $\phi_{\vec{k}}$ in the usual way:
\begin{equation} \label{phik}
\phi_{\vec{k}} = \psi_k(t)  A_{\vec{k}} + \psi^*_k(t)  A^\dagger_{\vec{k}} .
\end{equation}
Because of \eqref{fourier} and \eqref{evo}, the mode functions $\psi_k(t)$ satisfy
\begin{equation} \label{modes}
\ddot \psi_k + 3 H \dot \psi_k + k^2  \psi_k = 0 .
\end{equation}
In the Heisenberg picture, the time evolution of $\phi$ is encoded in the functions $\psi_k(t)$ and ${ A}_{\vec{k}}$ always annihilate the vacuum.  

The commutator between ${ A}$ and ${ A}^\dagger$ is proportional to the total volume
and therefore will receive the postulated corrections \eqref{2}. We make the following ansatz:
\begin{equation}  \label{piazza}
[{ A}_{\vec{k}},{ A}^\dagger_{\vec{k}'}] = (2 \pi L)^3 \ \delta_{{\vec{k}}, {\vec{k}'}}\ \left(
1 - \gamma \frac{H^2}{k_{\rm phys}^2} + {\cal O} (H L)^4 \right) \, .
\end{equation}
Since $k>1/L$ (the zero mode will not be considered), the correction is of the required type. The parameter $\gamma$ will be determined by applying the Ultra-Strong equivalence principle to the quadratically divergent part of the stress-energy tensor VEV.

It is convenient to consider the evolution with respect to conformal time $a d \tau = d t$. For a Universe with arbitrary equation of state $w = -(2\nu + 3)/(6\nu -3)$, the scale factor grows as $a(\tau) \propto \tau^{1/2 -\nu}$ and 
\begin{equation} \label{h}
H(\tau) a(\tau) = \frac{1 - 2\nu}{ 2 \tau}.
\end{equation}
 The solutions of \eqref{modes} are expressed in terms of Hankel functions of first and second type. The Bunch-Davis vacuum corresponds to the choice 
 \begin{equation} \label{hankel}
 \psi_k(\tau) = \tau ^\nu H_\nu^{(1)}(k \tau), \qquad  \psi^*_k(\tau) = \tau ^\nu H_\nu^{(2)}(k \tau).
 \end{equation}
We now want to calculate the vacuum expectation value of the energy density \eqref{hamtime} at $\vec{x} \approx 0$. Apart from order one factors, one obtains \cite{fulling}: 
\begin{equation} \label{t00}
\langle 0 |\, T_0^0 (\vec{x}\approx 0) \, | 0 \rangle =  \frac{a(\tau)^3}{(2 \pi L_0)^6} \sum_{\vec{k}, \vec{k}'}
\left(\dot{\psi}_k \dot{\psi}^*_{k'} - \frac{\vec{k} \cdot \vec{k}'}{a(\tau)^2}\, \psi_{\vec{k}} \psi_{\vec{k}'}\right)[A_{\vec{k}}\, ,\,  A_{-\vec{k}'}^\dagger].
\end{equation}
By plugging \eqref{hankel} and \eqref{piazza} in \eqref{t00},  and expanding at high $k$ we have:
\begin{equation} \label{orderbyorder}
\langle 0 |\, T_0^0 (0) \, | 0 \rangle \propto 
\sum_{\vec{k}} \left(\frac{k}{a}\right)\left(4 + \frac{(2\nu -1)^2}{2 (k  \tau)^2} +
{\cal O}(k \tau)^{-4} \right)\left(1 - \gamma \frac{(2\nu -1)^2}{4 (k  \tau)^2}\right)
\end{equation}
where the last factor comes from the anomalous  commutation relations \eqref{piazza} since, by \eqref{h}, $H^2/k_{\rm phys}^2 = (2\nu - 1)^2/4 k^2 \tau^2$. Note that the quadratic divergence can be reabsorbed, for every equation of state, by setting $\gamma = 1/2$. This is an encouraging result. 

A closer look at this model shows \cite{newfedo} that a compact Universe is in fact inconsistent with the present framework and that the Universe needs to be infinitely extended in every direction. The considered toy-Universe, however, helped fixing the ideas and find the correct recipe, which is encoded in the modified commutation relations \eqref{piazza} with $\gamma = 1/2$:
\begin{equation}  \label{piazza2}
[{ A}_{\vec{k}},{ A}^\dagger_{\vec{k}'}] = \delta^3( {\vec{k}} -  {\vec{k}'})\ \left(
1 - \frac{H^2 a^2}{2 k^2} \right) \, .
\end{equation}
Although in the text we generically refer to modifications at scales of the inverse curvature, we find, more precisely, that it is the extrinsic curvature which appears to regulate the correct modification.
By a ($k$- and $t$- dependent) rescaling of the $A$ operators one can recover the usual commutation relations. In fact,  the model can be equivalently formulated by keeping the usual ``metric manifold" operators and modifying in the IR the dispersion relations \cite{newfedo}.

\section{Discussion}

It is a well established paradigm that gravity and the matter fields are described by the theory
\begin{equation} \label{action}
S = \int \sqrt{-g}\left(M_{\rm Pl}^2 R + {\cal L}_{\rm matter}\right).
\end{equation}
and that such a theory can be trusted at least up to its semi-classical limit or at the low-energy effective level. The only problem with \eqref{action} is generally believed to be its UV-completion, for which many interesting alternatives (most notably, string theory) have been explored and studied. However, there are some known difficulties with \eqref{action}  which seem to be totally UV-insensitive, as they show up already at the effective/semi-classical level: 
\begin{itemize}
\item The cosmological constant is overestimated  by at least several orders of magnitude. 
\item The black hole information loss paradox: the quantized fields on a black hole background produce particles that radiate to infinity  (Hawking radiation). The black hole eventually evaporates and an initially pure state evolves into a thermal mixed state. 
\item There are models, within the framework \eqref{action}, that can describe our cosmological observations with impressive accuracy. However, from a theoretical point of view, those explanations are not without cost: we need to assume two epochs of accelerated expansion (inflation and ``dark energy") that require appropriate negative pressure components and, expecially for dark energy, a tremendous amount of fine tuning. 
\end{itemize}
Not surprisingly, the candidate theories of quantum gravity do not seem to be able to give a definite solution to any of the above. 

In this note we focused our attention to the IR side of the established picture and tried to look at \eqref{action} as a small-distance approximation. Instead of introducing a new (extremely small) mass scale, we find very compelling the possibility that modifications of \eqref{action} might appear at distances of the order of the curvature. We have tried to mimic the notable example of GR, that, with respect to Newtonian gravity, does not contain any more mass parameter but does constitute a substantially new theoretical framework.

The proposed modification is subject to the constraint of satisfying the ``Ultra-Strong" equivalence principle. The general motivations were given in Sec. \ref{secinv}. Beside, postulating the absence of any time-dependent term in the stress-energy tensor VEV appears to address, although quite drastically, the first two problems mentioned at the beginning of this section. By expanding around a flat FRW Universe, we found the first IR correction  \eqref{piazza2} of the field operators that reabsorbs the quadratic divergence of the stress-energy tensor VEV.  A more complete look at the global picture that emerges and at its cosmological consequences will be given elsewhere \cite{newfedo}.

{\bf Acknowledgments}
I thank Sergio Cacciatori, Fabio Costa, Olaf Dreyer, Filippo Passerini Costantinos Skordis and Andrew Tolley. Research at the Perimeter Institute is supported in part by the Government of Canada through NSERC and by the Province of Ontario through the Ministry of Research \& Innovation.\\

\end{document}